\documentclass[]{aa}  

\usepackage{graphicx}

\usepackage{amssymb}
\usepackage{amsmath}
\usepackage{color}
\usepackage{multirow}
\usepackage{makecell}
\usepackage{subcaption}
\usepackage{appendix}
\usepackage{caption}
\usepackage{tabularx}
\usepackage{graphicx,url,twoopt,natbib}
\usepackage[varg]{txfonts}
\usepackage{enumerate}
\usepackage[switch]{lineno}
\usepackage{ulem,soul}
\usepackage{threeparttable}
\usepackage{CJKutf8}
\bibpunct{(}{)}{;}{a}{}{,} 
\usepackage[colorlinks,
            linkcolor=red,
            anchorcolor=blue,
            citecolor=blue]{hyperref}

\newcommand{\mh}{[Fe/H]}

\begin{document} 
\begin{CJK}{UTF8}{gbsn}
   \title{Surface brightness–color relations for red giant branch stars: Observational constraints on metallicity effects using the ARD method}
   
   \author{Jianping Xiong$^{\dag}$
          \inst{1,2},
          Xiaodian Chen
          \inst{3,4},
          Jiao Li
          \inst{1,2},
          Qiyuan Cheng
          \inst{1,2,4},
          Xiaobin Zhang
          \inst{3,4},
          Chao Liu,
          \inst{5,4},
          Zhanwen Han
          \inst{1,2,4},
          \and
          Xuefei Chen$^{\dag}$\inst{1,2,4}
          }

   \institute{International Centre of Supernovae （ICESUN）, Yunnan Key Laboratory of Supernova Research, Yunnan Observatories, Chinese Academy of Sciences (CAS), Kunming 650216, China\\
              \email{cxf@ynao.ac.cn, xiongjianping@ynao.ac.cn}
         \and
             Key Laboratory for Structure and Evolution of Celestial Objects, Chinese Academy of Sciences, P.O. Box 110, Kunming 650216, People's Republic of China
        \and
            CAS Key Laboratory of Optical Astronomy, National Astronomical Observatories, Chinese Academy of Sciences, Beijing 100101, China 
        \and
            School of Astronomy and Space Science, University of the Chinese Academy of Sciences, Beijing 101408, China
        \and
            Key Laboratory of Space Astronomy and Technology, National Astronomical Observatories, Chinese Academy of Sciences, Beijing, 100101, China
             }

   \date{Received: \,\,\,\,\,\,\,\,accepted  }

 
\abstract
{The surface brightness-color relation (SBCR) links stellar color to angular diameter and is a key ingredient of geometric distance measurements on the first rung of the cosmic distance ladder. However, previous RGB-based calibrations were limited by small samples and did not observeably constrain the role of metallicity.}
{We aim to quantify the metallicity dependence of the SBCR for red giant branch (RGB) stars and to test the robustness of the relation using asteroseismic radii, Gaia distances, and atmospheric parameters from APOGEE.}
{We selected more than 2,000 RGB stars from APOKASC-3 to calibrate and validate the SBCR. Johnson $V$ magnitudes were synthesized from Gaia XP spectra and homogenized to widely used SBCR photometric systems, while $K_{\rm s}$ photometry was taken from 2MASS. Angular diameters derived from asteroseismic radii and Gaia distances (ARD) were used to construct the SBCR. We explored three fitting strategies: metallicity-free, metallicity-binned, and global metallicity-dependent relations.}
{Over the range $(V-K_{\rm s}) \approx 2$-3, the SBCR shows only a weak metallicity dependence. A change of 1 dex in [Fe/H] modifies the predicted angular diameter by less than 1\%, well below the intrinsic scatter of the calibration ($\sim$0.05 mag). This result is consistent with theoretical expectations. Comparison with the interferometric sample reveals a systematic offset of $\sim$1.5\% toward smaller angular diameters in our SBCR predictions, with a mild color dependence.
}
{The metallicity effect on the SBCR is small in the color range explored here, but it becomes relevant for sub-percent distance measurements. Our results show that large RGB samples with asteroseismic radii and Gaia distances provide a powerful observational route for SBCR calibration, with clear potential for extension to cooler and redder giants as the precision and parameter coverage of the input data improve.}

   \keywords{Stars: late-type --
             Asteroseismology --
             Parallaxes --
             Stars: fundamental parameters --
             Stars: distance
               }
\titlerunning{SBCR for RGB Stars using Asteroseismic radii and Gaia DR3}
\authorrunning{Jianping Xiong et al.}

\maketitle
%
\section{Introduction}
Accurate distance determination is fundamental to modern astrophysics, underpinning our understanding of stellar and galactic properties, as well as the construction of the cosmic distance ladder and precision cosmology. The surface brightness–color relation (SBCR), which links stellar color to angular diameter, plays a central role in this context. 
It is widely used to estimate stellar angular diameters from broad-band photometry 
\citep[e.g.][]{1976MNRAS.174..503B,1976MNRAS.174..489B,1997A&A...320..799F,2004A&A...426..297K}. In particular, it is a key ingredient in geometric distance measurements of late-type eclipsing binaries in the Magellanic Clouds 
\citep{2013Natur.495...76P,2019Natur.567..200P,2014ApJ...780...59G,2020ApJ...904...13G}. 
The resulting eclipsing-binary distances, especially the LMC distance, provide geometric anchors for the Cepheid-Type Ia supernova distance ladder and thus contribute to determinations of the Hubble constant \citep{2019ApJ...876...85R,2022ApJ...934L...7R}.
 SBCR is also used in Baade-Wesselink or infrared surface-brightness studies of Cepheids and RR Lyrae stars to infer angular-diameter variations from color variations 
\citep[e.g.][]{1997A&A...320..799F,2004A&A...428..587K,2011A&A...534A..94S,2024A&A...690A.295Z}. 
Therefore, a precise SBCR calibration is important for controlling systematic errors in these distance measurements.

The SBCR has traditionally been calibrated using angular diameters measured with long-baseline interferometry (LBI) \citep{2004A&A...426..297K,2005MNRAS.357..174D,2019Natur.567..200P,2020A&A...640A...2S}, enabling distance measurements to nearby galaxies with a precision of $\sim$1–2\% \citep{2019Natur.567..200P,2020ApJ...904...13G}. In the Gaia era, new independent calibration routes have emerged, including detached eclipsing binaries combined with parallaxes \citep{2021A&A...649A.109G}, and more recently, SBCRs based on asteroseismic radii and Gaia distances (ARD) \citep{2025AA...697A..37X}. While these developments have significantly improved statistical precision, systematic uncertainties are now becoming the dominant limitation. Discrepancies among different SBCR calibrations have been reported \citep{2020A&A...640A...2S}, likely arising from differences in angular-diameter measurements, photometric systems, extinction corrections, and stellar metallicity.

Among these sources, metallicity is of particular interest from a physical point of view. By modifying atmospheric opacity and temperature structure, metallicity can affect both stellar colors through line blanketing and the surface brightness itself via radiative transfer processes. Theoretical calculations based on MARCS model atmospheres predict a measurable, though generally weak, metallicity dependence of the SBCR for giants, with a stronger effect at redder colors \citep{2022A&A...662A.120S}. However, observational constraints on this dependence remain limited, largely due to the restricted size and parameter coverage of previous samples. This issue is particularly relevant for distance-scale applications, where mismatches between the metallicity of calibration samples and target populations may introduce systematic biases.

The availability of large samples of evolved stars with well-determined parameters now provides an opportunity to address this problem. Asteroseismic observations from missions such as Kepler and TESS have enabled the determination of stellar radii with typical precisions of $\sim$1\% \citep{2018ApJS..239...32P,2022ApJ...927..167L,2023ApJ...953..182W,2024AJ....167...50S,2025ApJS..276...69P}. Combined with Gaia DR3 parallaxes and distances \citep{2021AJ....161..147B,2023A&A...674A...1G}, these data allow robust angular-diameter estimates independent of interferometry. In addition, large spectroscopic surveys such as APOGEE and LAMOST provide homogeneous atmospheric parameters, including metallicity, while near-infrared photometry from 2MASS enables the construction of optical–infrared colors commonly used in SBCR studies. Building on our previous ARD calibration \citep{2025AA...697A..37X}, these datasets enable a systematic observational investigation of the metallicity dependence of the SBCR.

In this paper, we use a large sample of RGB stars with asteroseismic radii and Gaia DR3 distances to quantify the metallicity dependence of the SBCR and to place constraints on its impact. The structure of the paper is as follows. Section~\ref{data} describes sample selection and data reduction. Section~\ref{method} presents the methodology used to construct and analyze the SBCR. Section~\ref{result} presents the main results, including validation of the relation and its dependence on metallicity. Finally, Section~\ref{summary} discusses the implications of our findings and summarizes our conclusions.

\section{Data} \label{data}

In this paper, we use red giant stars from the APOKASC-3 catalog \citep{2025ApJS..276...69P}, which provides a homogeneous set of APOGEE spectroscopic parameters and \textit{Kepler} asteroseismic measurements for 15,808 evolved stars, including both red giant branch (RGB) and red clump (RC) stars. \citet{2025ApJS..276...69P} showed that asteroseismic radii are in best agreement with Gaia-based radii in the lower RGB, approximately corresponding to $\nu_{\rm max} \gtrsim 30\,\mu{\rm Hz}$. In this regime, the asteroseismic scaling relations are also most reliable, while larger systematic deviations may occur for more evolved giants or RC stars \citep{2013A&A...550A.126M,2016ApJ...822...15S}. 

We therefore restrict our analysis to RGB stars in this regime, resulting in an initial sample of about 4000 stars. Our RGB sample ($T_{\rm eff}\simeq4300$-5100 $\rm K$, $\log g\simeq2.1$-3.4 $\rm {dex}$) falls in the F5/K7-II/III giant category defined by \citet{2020A&A...640A...2S}. For these targets, we compile multi-band photometry and distance information, and apply a series of quality-control criteria as described below.

\subsection{Photometric measurements}

\subsubsection{Johnson $\rm V$-band magnitudes}

Accurate SBCR calibration requires high-precision Johnson $\rm V$-band photometry. However, large-scale surveys rarely provide homogeneous $V$-band measurements for large stellar samples. Existing catalogs, such as APASS\footnote{\url{https://www.astroplanner.net/apass.html}} (American Association of Variable Star Observers Photometric All-Sky Survey) and the All-Sky Compiled Catalog \citep{2009yCat.1280....0K}, suffer from limited coverage and heterogeneous photometric systems, which would significantly reduce the usable sample size. Gaia DR3 provides flux-calibrated low-resolution XP spectrophotometry for a large number of sources, covering a wavelength range of approximately 330–1050\,nm \citep{2023A&A...674A..33G,2023A&A...674A...2D}. This wide wavelength coverage makes it well suited for deriving synthetic broad-band magnitudes.
\begin{figure}[h]
    \centering
    \includegraphics[scale=0.75]{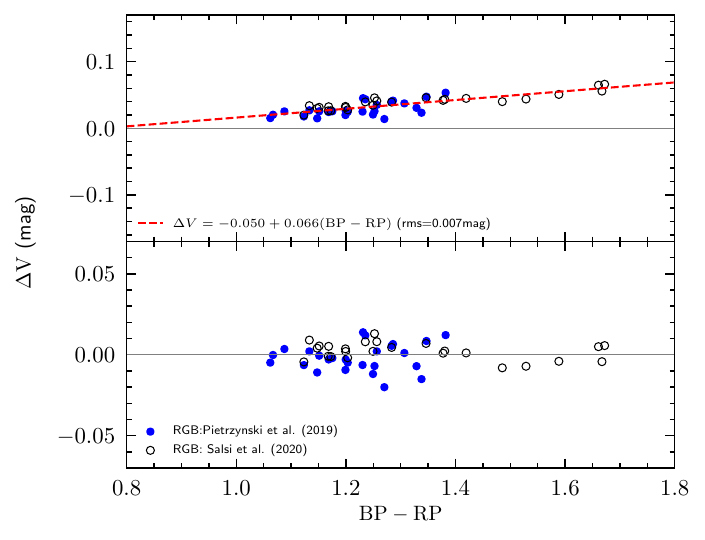}
    \caption{
Comparison between Gaia XP synthetic Johnson $V$ magnitudes and literature $V$ photometry for RGB stars from \citet{2019Natur.567..200P} and \citet{2020A&A...640A...2S}. 
The top panel shows the difference $\Delta V \equiv V_{\rm XP}-V_{\rm lit}$ as a function of the Gaia color index $(\mathrm{BP-RP})$. 
The bottom panel shows the residuals after applying this color correction. 
Blue filled circles represent RGB stars from \citet{2019Natur.567..200P}, and black open circles represent those from \citet{2020A&A...640A...2S}.
} \label{fig:Vmag_compare}
\end{figure} 

Therefore, in this work, we derive Johnson $\rm V$ magnitudes by convolving Gaia XP spectra with the \textit{GCPD/Johnson.V\_Landolt} transmission curve\footnote{\url{https://svo2.cab.inta-csic.es/svo/theory/fps3/index.php}}. This approach allows us to place our photometry on a system closely matched to that used in previous SBCR studies. Fig.~\ref{fig:Vmag_compare} compares our Gaia XP synthetic $V$ magnitudes ($V_{\rm XP}$) with literature measurements ($V_{\rm lit}$) from \citet{2019Natur.567..200P} and \citet{2020A&A...640A...2S}. A clear linear trend is observed between $\Delta V \equiv V_{\rm XP}-V_{\rm lit}$ and the Gaia color index $(\rm BP-RP)$, indicating a color-dependent offset between the two systems.

Such offsets arise naturally from differences in effective wavelength, bandpass shape, and flux calibration between Gaia XP-derived magnitudes and the standard Johnson system \citep{2005ARA&A..43..293B,2023A&A...674A...3M,2023A&A...674A...2D}. To account for this effect, we fit a linear relation as a function of $(\rm BP-RP)$ and use it to correct the synthetic $V$ magnitudes. After applying this correction, the color-dependent trend is largely removed and the residual scatter is significantly reduced (see the bottom panel in Fig.~\ref{fig:Vmag_compare}). The same correction is then applied to all RGB stars in our asteroseismic sample, this procedure ensures that the $V$-band photometry used in this work is homogeneous and directly comparable to previous SBCR calibrations.

\subsubsection{Infrared magnitudes}

The near-infrared photometry used in this work is taken from the Two Micron All Sky Survey (2MASS; \citealt{2006AJ....131.1163S}), which provides uniform all-sky observations in the $J$, $H$, and $K_{\rm s}$ bands. Owing to its homogeneous photometric system and high internal consistency, 2MASS has become the standard source of infrared magnitudes in SBCR studies. In particular, the $K_{\rm s}$ band is widely adopted because it is less affected by interstellar extinction and stellar atmospheric effects, making it well suited for optical–infrared SBCRs.

In this work, we adopt the $K_{\rm s}$ magnitudes from 2MASS and apply strict photometric quality criteria. We require the photometric quality flag to be \texttt{AAA}, corresponding to the highest-quality measurements in all three bands and a signal-to-noise ratio greater than 10 \citep{2006AJ....131.1163S}. In addition, we impose a constraint on the $K_{\rm s}$-band uncertainty of $\sigma_{K_{s}} <0.1$ \texttt{mag}. These criteria ensure that only high-quality infrared photometry is used in the SBCR construction.

\subsection{Distance and extinction corrections}

Distances are derived from Gaia DR3. To ensure reliable astrometric solutions, we exclude stars flagged as photometric variables, remove sources classified as non-single stars, and discard duplicated entries. We further require the Gaia renormalized unit weight error to satisfy $\mathrm{RUWE} \leq 1.4$, a commonly adopted criterion for selecting sources with well-behaved astrometry \citep{2021A&A...649A...2L,2018A&A...616A..17A}.

\begin{figure}[h]
    \centering
    \includegraphics[scale=0.8]{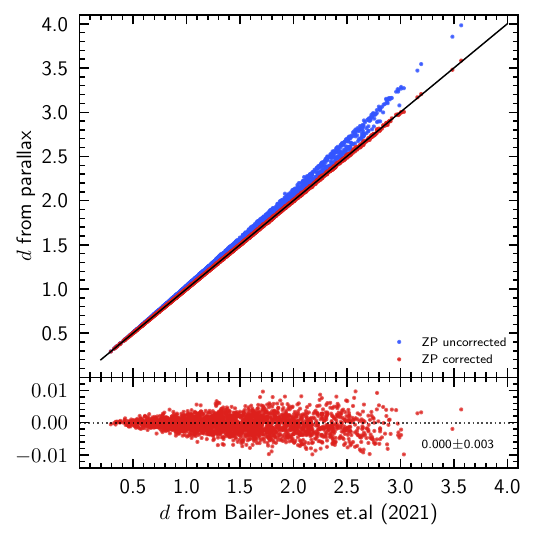}
    \caption{Comparison of Gaia-based distance estimates obtained using different methods. The $x$ axis shows the distances from \citet{2021AJ....161..147B}, while the $y$ axis shows distances derived from Gaia parallaxes. The blue points represent distances computed without correcting for the parallax zero-point offset, and the red points correspond to distances after applying the zero-point correction \citep{2021A&A...649A...4L}. The bottom panel displays the relative differences between the zero-point-corrected parallax distances and the distances from \citet{2021AJ....161..147B}.}\label{fig:distance_B_and_L}
\end{figure}

Distances are primarily adopted from Bayesian estimates of \citet{2021AJ....161..147B}. As a consistency check, we also compute distances using Gaia parallaxes corrected for the zero-point offset following \citet{2021A&A...649A...4L}. As shown in Fig.~\ref{fig:distance_B_and_L}, the two distance estimates are in good agreement, with no significant systematic offset and a dispersion of only $\sim$0.3\%. This indicates that the choice of distance estimator has a negligible impact on the derived angular diameters and SBCR calibration. We therefore adopt the distances from \citet{2021AJ....161..147B} in the following analysis.

For extinction corrections, we use the all-sky three-dimensional dust extinction map \footnote{\url{https://nadc.china-vo.org/data/dustmaps/}} of \citet{2025ApJS..280...15W}, which provides distance-dependent extinction estimates for individual stars. We adopt a standard extinction law with $R_V = 3.1$, and convert visual extinction to the infrared band using $A_{K_{\rm s}} = 0.078\,A_V$ \citep{2019ApJ...877..116W}.

After applying the above selection criteria, we exclude a small number of metal-poor stars with $\mathrm{[Fe/H]} < -1$, as their metallicity estimates tend to have larger uncertainties and could introduce additional scatter into the calibration. We also remove a few objects with extreme colors ($V-K_{\rm s} < 2$ or $V-K_{\rm s} > 3$), where the sample size is insufficient to robustly constrain the SBCR.

Finally, we obtain a total sample of $\sim$2400 RGB stars. The fitting subsample, consisting of $\sim$430 stars, is selected by requiring a 2\% precision in the angular diameter. The remaining stars form an independent validation sample that is used to assess the robustness of the derived SBCR.

\section{Method} \label{method}

\subsection{Surface brightness estimation}

The surface brightness of a star is defined as the flux emitted per unit angular area and can be expressed in terms of its intrinsic, extinction-corrected visual magnitude and limb-darkened angular diameter. Following \citet{1969MNRAS.144..297W}, the surface brightness in the $V$ band is given by

\begin{equation}
    S_{V} = V_{0} + 5 \log \theta_{LD} \label{Surface}
\end{equation}

where $S_{V}$ is the surface brightness in the $V$ band, $V_{0}$ is the extinction-corrected visual magnitude, and $\theta_{LD}$ is the limb-darkened angular diameter. In this work, the angular diameter is estimated as

\begin{equation} \label{theta}
    \theta_{LD} = \frac{2R}{d},
\end{equation}

where $R$ is the stellar radius derived from asteroseismology and $d$ is the distance inferred from Gaia \citep{2021AJ....161..147B}.

We compute $S_{V}$ using Eqs.~\ref{Surface} and \ref{theta}. For each star, the input quantities (photometry, distance, radius, and extinction) and their associated uncertainties are assumed to follow Gaussian distributions centered on the observed values. We then perform 2000 Monte Carlo realizations per target to propagate these uncertainties into the derived parameters. The final uncertainties are taken as the standard deviations of the resulting distributions.

\begin{figure}[h]
    \centering
    \includegraphics[scale=0.75]{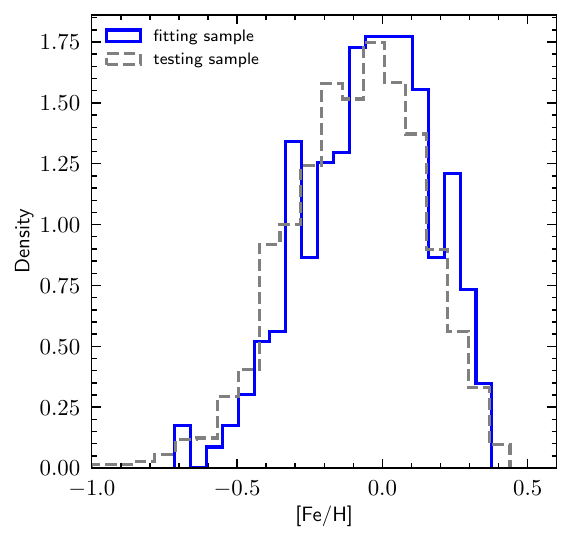}
    \caption{Distribution of \mh for the RGB stars used in this study. The blue solid line shows the fitting sample used to calibrate the SBCR, while the grey dashed line represents the testing sample. The two samples exhibit similar metallicity distributions.} \label{fig:feh_distribution}
\end{figure}

Fig.~\ref{fig:feh_distribution} shows the metallicity distributions of the fitting and validation samples. The metallicities are mainly distributed over $-0.5\texttt{dex} \lesssim \mathrm{[Fe/H]} \lesssim 0.5\texttt{dex}$, providing adequate coverage to investigate the metallicity dependence of the SBCR. The typical uncertainty in $S_{V}$ is about 0.03~\texttt{mag}.

\begin{figure}[h]
  \centering
   \includegraphics[scale=0.75]{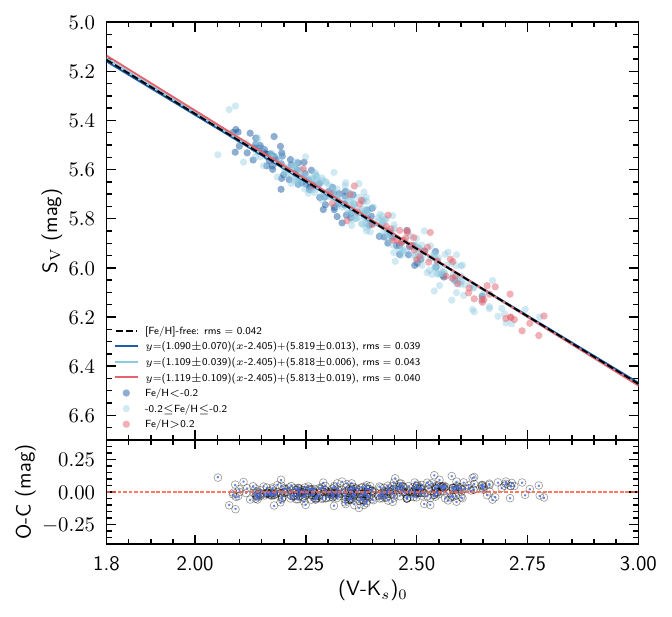}
  \centering
  \caption{Surface brightness–color relations fitted under different metallicity treatments. The dark blue, light blue, and red points represent RGB stars with $\mathrm{[Fe/H]} < -0.2$, $-0.2 \leq \mathrm{[Fe/H]} \leq 0.2$, and $\mathrm{[Fe/H]} > 0.2$, respectively. The dark blue, light blue, and red solid lines indicate the SBCR fits obtained in each metallicity bin, while the black dashed line shows the fit derived from the full sample without including a metallicity term (\texttt{[Fe/H]-free}). The bottom panel presents the residuals (O$-$C) of the fits, where the blue dots correspond to the metallicity-binned relations and the black open circles to the metallicity-free relation.} \label{fig:sbcr_fit}
\end{figure}

\subsection{SBCR fitting}
We adopt the following functional form for the surface brightness–color relation:

\begin{equation}
S_{V} = a \, \big[(V-K_{s})_{0} - 2.405\big] + c , \label{fitting_eq}
\end{equation}

where $(V-K_{s})_{0}$ is the extinction-corrected color index. The pivot value of 2.405 is adopted from \citet{2019Natur.567..200P}, as our color range closely matches that of their sample and this choice facilitates a direct comparison between different calibrations. The fitting is performed using a Markov Chain Monte Carlo (MCMC) approach, allowing us to estimate the model parameters and their uncertainties while properly accounting for observational errors.

To investigate the metallicity dependence of the SBCR, we consider three complementary fitting strategies. First, we divide the sample into three metallicity bins: $\mathrm{[Fe/H]} < -0.2$, $-0.2 \leq \mathrm{[Fe/H]} \leq 0.2$, and $\mathrm{[Fe/H]} > 0.2$, and independently fit Eq.~\ref{fitting_eq} to each subsample, allowing both the slope $a$ and intercept $c$ to vary. This approach enables us to examine possible systematic variations of the SBCR across different metallicity regimes.

Second, we include metallicity explicitly in a global fit by adopting

\begin{equation}
S_{V} = a \, \big[(V-K_{s})_{0} - 2.405\big] + b\,\mathrm{[Fe/H]} + c , \label{feh_global}
\end{equation}

where the coefficient $b$ quantifies the linear dependence of the SBCR on metallicity.

Finally, for comparison, we perform a global fit using the entire sample without including metallicity as an additional parameter, representing the conventional SBCR calibration that neglects metallicity effects.

These three approaches provide a consistent framework to assess both the significance and the form of the metallicity dependence. The resulting relations and their implications are presented and discussed in Sect.~\ref{result}.

\begin{figure}[h]
  \centering
   \includegraphics[scale=0.8]{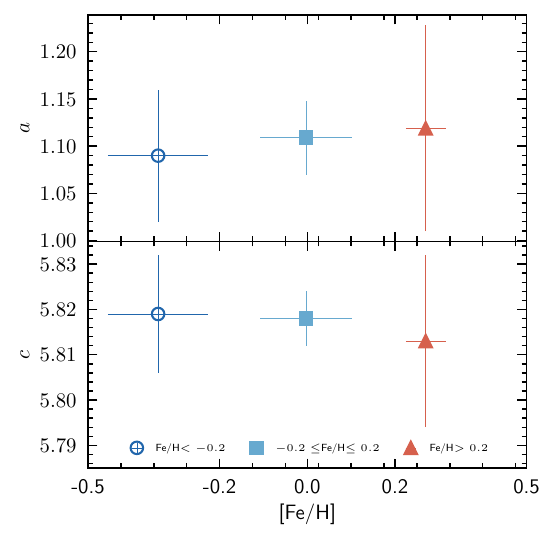}
  \centering
  \caption{Variation of the SBCR fitting parameters $a$ and $c$ as a function of metallicity. The relation is parameterized as Eq.\ref{fitting_eq}. The dark blue circle, light blue square, and red triangle correspond to the metallicity bins $\mathrm{[Fe/H]} < -0.2$, $-0.2 \leq \mathrm{[Fe/H]} \leq 0.2$, and $\mathrm{[Fe/H]} > 0.2$, respectively. Error bars indicate the uncertainties of the fitted parameters.} \label{fig:slope}
\end{figure}

\begin{table*}[htbp]
\centering
\caption{Best-fit SBCR calibrations for different metallicity strategies and comparison with previous works.}
\label{tab:sbcr_fit}
\setlength{\tabcolsep}{4pt}
\renewcommand{\arraystretch}{1.10}

\begin{tabular}{@{}>{\footnotesize}l 
                >{\raggedright\arraybackslash}m{7.2cm} 
                >{\footnotesize}c 
                >{\footnotesize}c 
                >{\footnotesize}l@{}}
\hline\hline
Method & SBCR ($S_V$) & rms (mag) & $S_{V}$ at color $=2.5$ & Source \\
\hline

\multicolumn{5}{@{}l}{\textit{This work}} \\
\hline

\multicolumn{5}{@{}l}{[Fe/H]-binned} \\

\quad $\mathrm{[Fe/H]} < -0.2$ &
{\scriptsize $(1.090\pm0.070)\,\big[(V-K_{\rm s})_{0}-2.405\big]+(5.819\pm0.013)$} &
0.039 &
5.923 &
This work \\

\quad $-0.2 \leq \mathrm{[Fe/H]} \leq 0.2$ &
{\scriptsize $(1.109\pm0.039)\,\big[(V-K_{\rm s})_{0}-2.405\big]+(5.818\pm0.006)$} &
0.043 &
5.923 &
This work \\

\quad $\mathrm{[Fe/H]} > 0.2$ &
{\scriptsize $(1.119\pm0.109)\,\big[(V-K_{\rm s})_{0}-2.405\big]+(5.813\pm0.019)$} &
0.040 &
5.919 &
This work \\

[Fe/H]-free &
{\scriptsize $(1.099\pm0.029)\,\big[(V-K_{\rm s})_{0}-2.405\big]+(5.818\pm0.005)$} &
0.042 &
5.922 &
This work \\

[Fe/H]-global &
{\scriptsize $(1.106\pm0.033)\,\big[(V-K_{\rm s})_{0}-2.405\big]+(-0.010\pm0.026)\,\mathrm{[Fe/H]}+(5.818\pm0.005)$} &
0.042 &
\makecell{$5.918$ \scriptsize ([Fe/H]$=+0.5$)\\
          $5.923$ \scriptsize ([Fe/H]$=0.0$)\\
          $5.928$ \scriptsize ([Fe/H]$=-0.5$)} &
This work \\

\hline
\multicolumn{5}{@{}l}{\textit{Previous RGB and late-type giant SBCRs}} \\
\hline

RGB (LBI) &
{\scriptsize $(1.330\pm0.017)\,\big[(V-K)_{0}-2.405\big]+(5.869\pm0.003)$} &
0.018 &
5.995 &
\citet{2019Natur.567..200P} \\

RGB (ARD) &
{\scriptsize $(1.230\pm0.021)\,(V-K_{\rm s})_{0}+(2.845\pm0.048)$} &
0.075 &
5.920 &
\citet{2025AA...697A..37X} \\

F5/K7-II/III (LBI) &
{\scriptsize $1.220\,(V-K_{\rm s})_{0}+2.864$} &
0.002 &
5.914 &
\citet{2021AA...652A..26S} \\

\hline
\multicolumn{5}{@{}l}{\textit{Cepheid PARSEC relations}} \\
\hline

\quad $\mathrm{[M/H]}=0.0$ &
{\scriptsize $1.249\,(V-K)_{0}+2.784$} &
-- &
5.907 &
\citet{2025AA...698A..46B} \\

\quad $\mathrm{[M/H]}=-0.5$ &
{\scriptsize $1.241\,(V-K)_{0}+2.841$} &
-- &
5.943 &
\citet{2025AA...698A..46B} \\

\hline
\end{tabular}

\end{table*}

\section{Results} \label{result}

\subsection{SBCR for different \mh}

The fitting results are presented in Fig.~\ref{fig:sbcr_fit}, with the metallicity dependence of the SBCR parameters shown in Fig.~\ref{fig:slope}. In Fig.~\ref{fig:sbcr_fit}, the observed data (colored points) show a mild correlation between metallicity and color, with metal-rich stars systematically occupying redder $(V-K_{s})_{0}$ values, consistent with expectations from line-blanketing effects. At a given color, a small offset in surface brightness is also observed, in the sense that higher-metallicity stars tend to have slightly lower $S_{V}$ (e.g. around $(V-K_{s})_{0} \simeq 2.25$), although the amplitude of this effect remains limited. The corresponding best-fit relations for different metallicity bins (colored solid lines) closely follow the overall trend of the data, while the global fit without a metallicity term (black dashed line) provides a similarly good description. The residuals (O$-$C), shown in the bottom panel, are centered around zero with no significant dependence on color for both the metallicity-binned relations (blue dots) and the metallicity-free relation (black open circles). The dispersion of the relation is $\sim$0.046–0.050~ \texttt{mag}, indicating that the adopted linear formulation provides an adequate description of the data in the explored parameter space. No significant reduction in scatter is obtained when including metallicity, suggesting that its impact is minor within this color range.

Furthermore, the metallicity dependence of the fitted parameters is illustrated in Fig.~\ref{fig:slope}. The slope $a$ increases slightly from $1.101\pm0.048$ for $\mathrm{[Fe/H]}<-0.2$ to $1.256\pm0.086$ for $\mathrm{[Fe/H]}>0.2$, while the intercept $c$ shows a marginal decrease from $5.820\pm0.008$ to $5.802\pm0.017$. However, given the uncertainties, these variations are not statistically significant, and the \texttt{rms} scatter remains nearly unchanged across the metallicity bins.

The best-fit parameters are summarized in Table~\ref{tab:sbcr_fit}, together with comparisons to our previous ARD calibration \citep{2025AA...697A..37X}, as well as to the relations of \citet{2019Natur.567..200P} and \citet{2021AA...652A..26S}. For the color range explored in this work ($V-K_{\rm s}\approx2.0$–3), the relations derived in different metallicity bins yield very similar predictions. As shown in Table~\ref{tab:sbcr_fit}, at $(V-K_{\rm s})_{0}=2.5$, the difference in $S_{V}$ among the three bins is less than 0.01~\texttt{mag}, much smaller than the intrinsic scatter of the relation. For the metallicity-dependent global fit (Eq.~\ref{feh_global}), the predicted $S_{V}$ values at $\mathrm{[Fe/H]}=-0.5~\texttt{dex}$ and $+0.5~\texttt{dex}$ differ by about 0.01 \texttt{mag}, corresponding to $\sim$0.5\% in angular diameter. For a metallicity difference of 0.5~\texttt{dex}, more typical of nearby galaxies, the variation in $S_{V}$ is $<$0.01~\texttt{mag}. This weak metallicity dependence is consistent with the theoretical predictions of \citet{2022A&A...662A.120S}. Their MARCS-based calculations show that metallicity effects on the SBCR are strongly color dependent, remaining small over the color range $V-K_{\rm s}\approx2$–3 and becoming significant only toward redder colors. In particular, they find that the difference in inferred angular diameter between $\mathrm{[Fe/H]}=0.0$ and $-0.5$ is below $\sim$0.4\% in this regime, in good agreement with our results. In addition, a similar conclusion was recently obtained by \citet{2025AA...698A..46B} using PARSEC models for Cepheids (see Table~\ref{tab:sbcr_fit}). They investigated the metallicity dependence of Cepheid SBCRs over the range $-0.5 < [{\rm M/H}] < 0.5$ and found that the SBCR depends only weakly on metallicity. For example, at $V-K=2.5$, their PARSEC-based Cepheid models predict that a 0.5~dex change in $[{\rm M/H}]$ produces a variation of only $\sim0.03$~mag in $S_V$, corresponding to an angular-diameter change of $\sim1.4\%$.
\begin{figure*}[htbp]
\centering
   \includegraphics[scale=0.75]{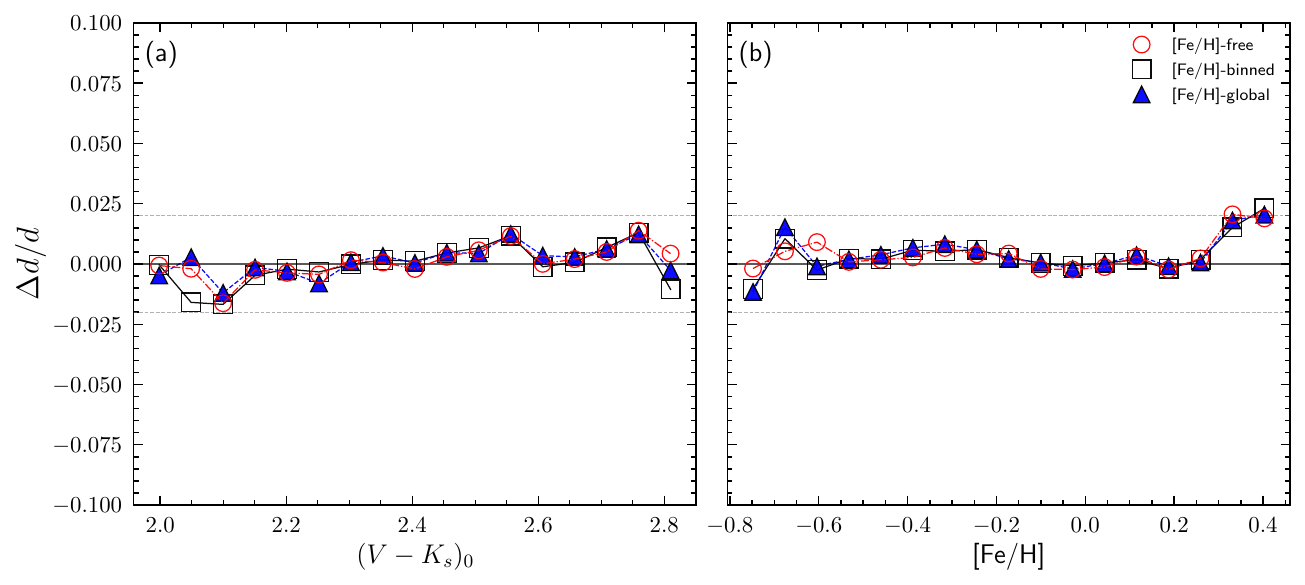}
  \centering
  \caption{
Comparison between SBCR-based distances ($d_{\rm SBCR}$) and Gaia distances ($d_{\rm Gaia}$) for the independent testing sample, shown as a function of color and metallicity. The red open circles, grey open squares, and blue filled triangles correspond to the \texttt{[Fe/H]-free}, \texttt{[Fe/H]-binned}, and \texttt{[Fe/H]-global} relations, respectively (see Table~\ref{tab:sbcr_fit} for definitions). The left panel shows the relative distance residuals as a function of $(V-K_s)_0$, while the right panel shows their dependence on $\mathrm{[Fe/H]}$. The vertical axis represents $(d_{\rm SBCR}-d_{\rm Gaia})/d_{\rm Gaia}$. The two dashed horizontal lines indicate $\pm2\%$ deviations. Each point corresponds to the mean value within bins of color or metallicity. Slightly larger deviations near the edges of the parameter space are likely due to statistical fluctuations and limited sampling in these regions.}
\label{fig:distance_comp}

\end{figure*}
\begin{figure}[h]
\centering
   \includegraphics[scale=0.75]{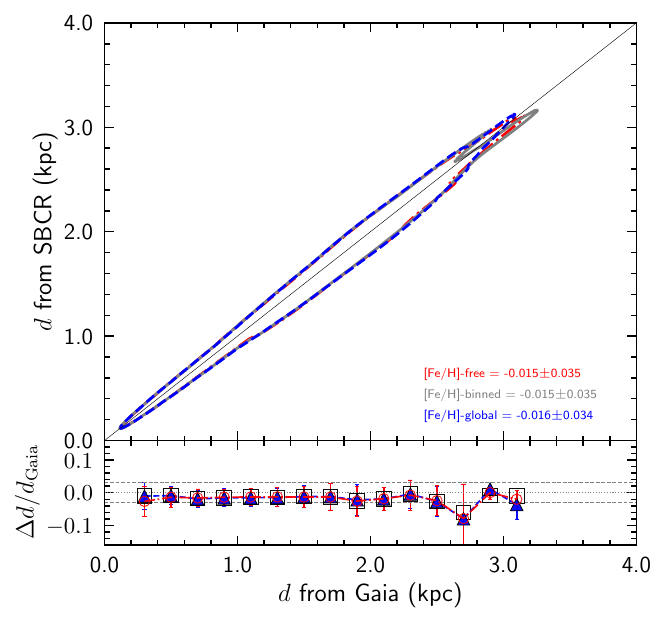}
  \centering
  \caption{Comparison between SBCR-based distances and Gaia distances using an independent sample of RGB stars with asteroseismic radii from \citet{2023ApJ...953..182W}. The upper panel compares the Gaia-based distances with the distances inferred from our SBCR. The outer Kernel density estimation (KDE) contours show the density distributions for the three SBCR prescriptions. The red dash-dotted, grey solid, and blue dashed contours correspond to the \texttt{[Fe/H]-free}, \texttt{[Fe/H]-binned}, and \texttt{[Fe/H]-global} relations, respectively (see Table~\ref{tab:sbcr_fit} for definitions). The lower panel presents the relative differences $(d_{\rm SBCR}-d_{\rm Gaia})/d_{\rm Gaia}$ in bins of distance. The red open circles, grey open squares, and blue filled triangles correspond to the \texttt{[Fe/H]-free}, \texttt{[Fe/H]-binned}, and \texttt{[Fe/H]-global} relations, respectively.} \label{fig:distance_comp_Wang}
\end{figure}

We note that \citet{2025AA...698A..46B} use $[{\rm M/H}]$, whereas our work and \citet{2022A&A...662A.120S} adopt $[{\rm Fe/H}]$. For a solar-scaled composition, $[{\rm M/H}]$ is approximately equivalent to $[{\rm Fe/H}]$; if $\alpha$-enhancement is present, the corresponding $[{\rm Fe/H}]$ would be lower than $[{\rm M/H}]$. Therefore, a 0.5~dex variation in $[{\rm M/H}]$ may correspond to a variation larger than 0.5~dex in $[{\rm Fe/H}]$, which could partly explain why the PARSEC-based Cepheid models predict a slightly larger change in $S_V$ than those inferred from the MARCS-based models and from our RGB-based ARD calibration. Nevertheless, this difference remains within the rms scatter of the SBCR. These comparisons indicate that the metallicity dependence inferred from our calibration is broadly consistent with both MARCS- and PARSEC-based theoretical predictions, all suggesting only a weak metallicity dependence of the SBCR over the color range considered here.

The comparison with the ARD calibration further shows that SBCRs based on asteroseismic radii and Gaia distances are broadly consistent. However, after placing our $V$-band photometry on a system consistent with that of \citet{2019Natur.567..200P}, we still find a systematic offset of about 0.07~\texttt{mag} between the relations, suggesting that the remaining discrepancy is likely related to differences in angular-diameter scales rather than photometric calibration.

\subsection{Comparison of distances}

To assess the robustness of our SBCR calibrations, we first compare the distances derived from different fitting strategies with Gaia-based distances, as shown in Fig.~\ref{fig:distance_comp}. Fig.~\ref{fig:distance_comp} includes both the fitting sample (top panels) and the independent testing sample (bottom panels). The red open circles, grey open squares, and blue filled triangles denote the results from the \texttt{[Fe/H]-free}, \texttt{[Fe/H]-binned}, and \texttt{[Fe/H]-global} relations, respectively. Gaia distances are adopted from the Bayesian estimates of \citet{2021AJ....161..147B}. The vertical axis shows the relative difference $(d_{\rm SBCR}-d_{\rm Gaia})/d_{\rm Gaia}$, and each point represents the mean value within bins of color or metallicity.

As shown in Fig.~\ref{fig:distance_comp}, for the independent testing samples, the residuals are centered around zero over the full ranges of $(V-K_{\rm s})_0$ and $\mathrm{[Fe/H]}$, with no evidence for systematic trends with either parameter. The three fitting strategies yield mutually consistent results within the uncertainties. Although slightly larger deviations are observed in a few bins near the boundaries of the parameter space, these remain within the overall dispersion and are likely driven by statistical fluctuations and edge effects, rather than indicating any systematic bias. For the independent testing sample, the scatter of $(d_{\rm SBCR}-d_{\rm Gaia})/d_{\rm Gaia}$ is at the level of $\sim$4\%, while the mean offset remains close to zero. This scatter is comparable to the intrinsic uncertainties of Gaia-based distances, indicating that the SBCR calibration does not introduce additional systematic dispersion. We note that the Gaia distances adopted here are taken from \citet{2021AJ....161..147B}. As shown in Sect.~\ref{data}, these Bayesian distances are in good agreement with distances derived from Gaia parallaxes corrected for the zero-point offset following \citet{2021A&A...649A...4L}, with no significant systematic offset and a dispersion of only $\sim$0.3\%. This consistency indicates that the adopted distance scale is internally coherent and does not introduce additional unaccounted systematic effects related to Gaia parallaxes.

We then perform an additional external test using the independent sample of red giants presented by \citet{2023ApJ...953..182W}, which provides high-precision asteroseismic radii derived with a different methodology. This sample allows us to examine whether our SBCR calibration remains stable when applied to radii obtained from an alternative asteroseismic pipeline. For these stars, the $V$-band magnitudes are synthesized from Gaia XP spectra and corrected following the procedure described in Sect.~\ref{data}, while the $K_{\rm s}$-band photometry is taken from 2MASS.

The comparison is shown in Fig.~\ref{fig:distance_comp_Wang}, where the horizontal axis gives the Gaia-based distances from \citet{2021AJ....161..147B}, and the vertical axis shows the distances inferred from our SBCR calibration. No significant systematic offset is found between the two distance estimates, and the scatter is at the level of $\sim$3\%, comparable to that obtained in the Gaia-based validation above. This indicates that our SBCR calibration is not sensitive to the specific choice of asteroseismic radius scale, at least within the precision of the current data.

Taken together, these tests show that introducing metallicity does not produce any measurable bias in the distance scale within the explored parameter space, and that our calibration yields distances fully consistent with Gaia for both the original sample and an independent set of stars with alternative asteroseismic radii. In the context of angular-diameter determination and application, this further indicates that the adopted distance term $d$ is internally consistent and does not reveal any additional unaccounted systematic bias associated with Gaia parallaxes. This supports the robustness and broad applicability of the present SBCR calibration.

\subsection{Comparison of angular diameters}
For an external validation of the derived SBCRs, we compiled a sample from \citet{2019Natur.567..200P,2020A&A...640A...2S}. We then excluded the M-II/III sample from \citet{2020A&A...640A...2S}, because M-type giants can be affected by strong molecular absorption features, which may alter the continuum, and because these cooler stars lie outside the $(V-K_{\rm s})_0$ color range covered by our APOKASC-3 RGB sample. The remaining stars have published limb-darkened angular diameters measured from long-baseline infrared interferometry, primarily based on ESO VLTI/PIONIER observations or compiled from the JMDC (JMMC Measured stellar Diameters Catalog; \citet{2016yCat.2345....0D}). These interferometric angular diameters typically reach a precision of about 1\% and are adopted here as reference measurements. For the comparison sample, we collected metallicity information from the \texttt{SIMBAD} database. We then applied the three fitting strategies (the \texttt{[Fe/H]-free}, \texttt{[Fe/H]-global}, and \texttt{[Fe/H]-binned} relations) developed in this work. Using the $V-K_{\rm s}$ color, we derived the SBCR-based limb-darkened angular diameter, $\theta_{\rm LD}$, from Eq.~\ref{Surface} through Eq.~\ref{theta_Sv}, and compared it with the interferometric measurements.

\begin{equation}
   \theta_{\rm{LD}} = 10^{0.2(S_{V}-V_{0})} \label{theta_Sv}
\end{equation} 

\begin{figure}[h]
  \centering
   \includegraphics[scale=0.75]{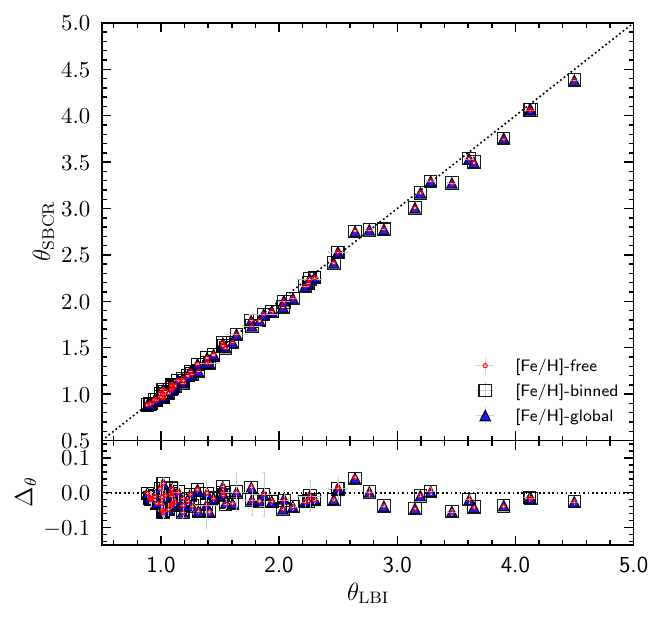}
  \centering
  \caption{Comparison between SBCR-predicted angular diameters and interferometric measurements. The $x$ axis shows angular diameters from long-baseline interferometry ($\theta_{\rm LBI}$), and the $y$ axis shows those inferred from our SBCR ($\theta_{\rm SBCR}$). Different symbols denote the three metallicity treatments (Table~\ref{tab:sbcr_fit}): \texttt{[Fe/H]-free} (red open circles), \texttt{[Fe/H]-binned} (grey open squares), and \texttt{[Fe/H]-global} (blue filled triangles). The lower panel shows the relative differences $(\theta_{\rm SBCR}-\theta_{\rm LBI})/\theta_{\rm LBI}$.} \label{fig:theta_compare}
\end{figure}

Fig.~\ref{fig:theta_compare} compares the angular diameters predicted from our SBCRs with the interferometric measurements. The $x$ axis shows the interferometric angular diameters ($\theta_{\rm{LBI}}$), while the $y$ axis gives the SBCR-based estimates ($\theta_{\rm{SBCR}}$). The bottom panel presents the relative residuals, defined as $(\theta_{\rm{SBCR}}-\theta_{\rm{LBI}})/\theta_{\rm{LBI}}$.

We find that the angular diameters derived from our SBCRs are systematically offset from the interferometric values by about $\sim$1.5\%. Such systematic differences between asteroseismic and interferometric scales have also been reported in previous studies, including our earlier work \citep{2025AA...697A..37X} and the analysis of \citet{2024A&A...690A.327V}. The results obtained from the \texttt{[Fe/H]-global}, \texttt{[Fe/H]-free}, and \texttt{[Fe/H]-binned} relations are highly consistent with each other, indicating that the choice of metallicity treatment has only a minor impact on the predicted angular diameters within this color range.

\begin{figure}[h]
  \centering
   \includegraphics[scale=0.85]{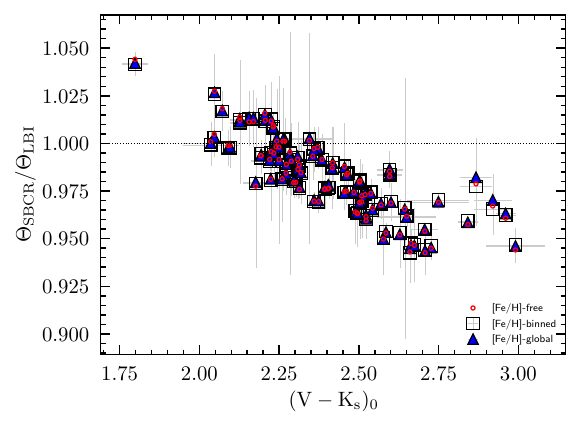}
  \centering
  \caption{Dependence of the ratio between SBCR-derived and interferometric angular diameters on color. The three metallicity treatments (Table~\ref{tab:sbcr_fit}) are color-coded as follows: red open circles (\texttt{[Fe/H]-free}), grey open squares (\texttt{[Fe/H]-binned}), and blue filled triangles (\texttt{[Fe/H]-global}). The dashed line marks $\theta_{\rm SBCR}/\theta_{\rm LBI}=1$.} \label{fig:theta_VK}
\end{figure}

In addition, we find a weak color dependence in the residuals. As shown in Fig.~\ref{fig:theta_VK}, the ratio $\theta_{\rm SBCR}/\theta_{\rm LBI}$ decreases slightly with increasing $(V-K)_{0}$. Larger $(V-K)_{0}$ values correspond to cooler and more evolved giants, where uncertainties in photometric calibration, limb-darkening corrections, and stellar atmosphere modeling are expected to become more significant, potentially contributing to the observed systematic trend. This behavior further suggests that residual systematics in current SBCR calibrations are more likely dominated by the underlying angular-diameter scale rather than by metallicity effects.

\section{Discussion and conclusions} \label{summary}

In this work, we provide an observational constraint on the metallicity sensitivity of the surface brightness-color relation using red giant stars with high-precision asteroseismic radii and \textit{Gaia}-based distances. Our calibration is based on more than 2000 RGB stars from APOKASC-3. The $V$ magnitudes derived from \textit{Gaia} XP spectra were carefully aligned with the photometric systems adopted in widely used SBCR calibrations \citep{2019Natur.567..200P,2020A&A...640A...2S} in order to minimize zero-point offsets. Within the $(V-K_{\rm s}) \simeq 2$--3 color range, we find a very weak metallicity dependence: a difference of 1 dex in [Fe/H] changes the predicted angular diameter by less than $\sim$1\%. This variation is smaller than the intrinsic scatter of our calibration (rms $\sim$0.04\,mag). At the same time, the intrinsic scatter is reduced from $\sim$0.075\,mag in previous ARD-based SBCRs \citep{2025AA...697A..37X}, reflecting the improved photometric consistency achieved with \textit{Gaia} XP-derived $V$ magnitudes.

The weak metallicity dependence of the SBCR implies that, for typical red giants, the effect of metallicity on distance measurements is small. For example, the Large Magellanic Cloud (LMC) has $\mathrm{[Fe/H]} \approx -0.4$\, \texttt{dex}, for which applying a solar-metallicity SBCR would shift the predicted angular diameters by less than 0.5\%. However, for geometric distance measurements aiming at sub-percent precision, even such small deviations cannot be neglected. Moreover, the metallicity-induced SBCR offset increases with $(V-K_s)$, implying that metallicity corrections will become increasingly important when extending SBCR-based distance determinations to redder and more luminous giants. Accounting for this effect is therefore relevant for high-precision calibration of the cosmic distance ladder and, ultimately, the Hubble constant.

Finally, after homogenizing the $V$-band photometry to match the interferometric system, we find a systematic offset of $\sim$1.5\% in angular diameter, with the SBCR-predicted values being slightly smaller than the interferometric measurements, as previously reported in the literature \citep{2024A&A...690A.327V,2025AA...697A..37X}. This offset also shows a mild color dependence, likely reflecting residual differences between asteroseismic and interferometric radius scales, as well as color-dependent uncertainties in photometric calibration, limb-darkening corrections, and stellar atmosphere models. In particular, asteroseismic radii depend explicitly on the effective temperature through the scaling relations, while interferometric limb-darkened angular diameters rely on model-dependent corrections, which may contribute to the observed systematic trend. In this context, obtaining a larger sample of well-characterized detached eclipsing binaries will be crucial for independently testing the SBCR zero point and disentangling remaining systematics in the angular-diameter scale. Extending the calibration to cooler giants and a broader color range, together with further improvements in asteroseismic radii, \textit{Gaia} distances, and independent eclipsing-binary constraints, will enable more precise SBCR calibrations and enhance their applicability to high-accuracy extragalactic distance measurements.

\section{Data availability}

The full dataset used in this work is available in electronic form at the the CDS via anonymous ftp to cdsarc.u-strasbg.fr (130.79.128.5) or via \url{http://cdsweb.u-strasbg.fr/cgi-bin/qcat?J/A+A/} (VizieR: to be assigned)

\begin{acknowledgements}
This work is supported by the National Natural Science Foundation of China (NSFC) with grant Nos.12125303, 12288102, 12090040/3, the National Key R\&D Program of China (grant No.2021YFA1600401/ 2021YFA1600403), the NSFC (grant Nos.12303106, 12473034, 12103086, 12373037, 12322306), Yunnan Fundamental Research Projects (grant Nos.202401AT070139, 202101AU070276), the Yunnan Revitalization Talent Support Program—Science \& Technology Champion Project (No.202305AB350003), the New Cornerstone Science Foundation through the XPLORER PRIZE, the CAS Project for Young Scientists in Basic Research (YSBR-148) and the International Centre of Supernovae, Yunnan Key Laboratory (No.202302AN360001). 

We thank the Gaia Data Processing and Analysis Consortium (DPAC) for their substantial contributions in producing and releasing high-quality data.
\end{acknowledgements}

\bibliographystyle{aa}
\bibliography{star}

\end{CJK}
\end{document}